\documentstyle{article}                                                 
\begin{document}     
\begin{center}{\Large\bf Five-Dimensional Brane World Theory}\end{center}
\begin{center}{Moshe Carmeli$^\star$}\end{center}
\begin{center}{Department of Physics, Ben Gurion University, Beer Sheva 84105, 
Israel}\end{center}
\renewcommand{\abstractname}{}
\begin{abstract}
A five-dimensional cosmological theory of gravitation that unifies space, time
and velocity is presented. Within the framework of this theory we first discuss some
general aspects of the universe in five dimensions. We then find the 
equations of motion of the expanding universe and show that it is accelerating.
This followed by dealing with the important problem of halo dark matter around
galaxies by deriving the equations of motion of a star moving around the 
field of a spherically-symmetric galaxy. The equations obtained are not Newtonian;
rather, the Tully-Fisher formula is obtained. The 
cosmological constant is subsequently discussed: our theory predicts that 
$\Lambda\approx 3H_0^2\approx 1.934\times 10^{-35}$s$^{-2}$, in agreement with
experimental results obtained by the High-Z Supernova Team and the Supernova
Cosmology Project. Finally we
derive a formula for the cosmological redshift in which appears the expression
($1-\Omega_M$), thus enabling us to determine the kind of the universe by means of
the cosmological redshift. We find that $\Omega_M$ should be less than 1 in
order not to contradict redshift measurements,
and therefore the universe is open.
\end{abstract}   
\begin{center}{PACS numbers: 04.50.+h, 11.25.Mj, 11.27.+d, 98.80.Cq}\end{center}
\vspace{2cm}
$^\star$  Email: carmelim@bgumail.bgu.ac.il
\newpage
\section{Introduction}
In this paper we present a five-dimensional cosmological theory of space, 
time and velocity. The added extra dimension of velocity to the usual 
four-dimensional spacetime will be evident in the sequel. Before introducing the
theory we have to deal, as usual, with coordinate systems in cosmology. Other
important basic issues will be dealt later on.  
\subsection{Cosmic Coordinate Systems: The Hubble Transformation}
We will use {\it cosmic coordinate systems} that fill up spacetime. Given one system $x$, 
there is another one $x'$ that differs from the original one by a {\it Hubble
transformation} 
$$x'=x+t_1v,\hspace{5mm} t_1=\mbox{\rm constant}, \eqno(1.1)$$
where $v$ is a velocity parameter, and $y$ and $z$ are kept unchanged. A third system will be given by another
Hubble transformation,
$$x''=x'+t_2v=x+(t_1+t_2)v.\eqno(1.2)$$ 

The cosmic coordinate systems are similar to the inertial coordinate systems,
but now the velocity parameter takes over the time parameter and visa versa.
The analogous Galileo transformation to Eq. (1.2) that relates inertial 
coordinate systems is given, 
as is known, by 
$$x''=x'+v_2t=x+(v_1+v_2)t.\eqno(1.3)$$

The universe expansion is also given by a formula of the above kind:
$$x'=x+\tau v,\eqno(1.4)$$
where $\tau=H_0^{-1}$ in the limit of zero distance, and thus a universal 
constant. (Its value is calculated in Subsection 5D as $\tau=12.486$Gyr.) 
However, the universe 
expansion is apparently incompatible with the Hubble spacetime transformation, 
namely one cannot add them. Thus, if we have 
$$x''=x'+tv, \hspace{5mm} x'=\tau v,\eqno(1.5)$$
then 
$$x''\neq\left(\tau+t\right)v.\eqno(1.6)$$
Rather, it is always
$$x''=\tau v.\eqno(1.7)$$
The above can be looked upon as a postulate of the theory.

This situation is like that we have with the propagation of light, 
$$x''\neq\left(c+v\right)t,\eqno(1.8)$$
but it is always
$$x''=ct\eqno(1.9)$$
in all inertial coordinate systems, and where $c$ is the speed of light in
vacuum.

The constancy of the speed of light and the validity of the laws of nature in
inertial coordinate systems, though they are both experimentally valid, they
are apparently not compatible with each other. We have the same situation in cosmology; 
the constancy of the Hubble constant in the zero-distance limit, and the
validity of the laws of nature in cosmic coordinate systems, though both are
valid, they are apparently incompatible with each other. 
\subsection{Lorentz-like Cosmological Transformation}
In the case of light propagation, one has to abandon the Galileo 
transformation in favor of the Lorentz transformation. In cosmology one has
to give up the Hubble transformation in favor of a new Lorentz-like 
cosmological transformation given by [1]
$$x'=\frac{x-tv}{\sqrt{1-t^2/\tau^2}},\hspace{5mm} v'=
\frac{v-tx/\tau^2}{\sqrt{1-t^2/\tau^2}},\hspace{5mm} y'=y, \hspace{5mm} z'=z,
\eqno(1.10)$$
for a motion with fixed $y$ and $z$. 

As is well known, the flat spacetime line element in special relativity is 
given by 
$$ds^2=c^2dt^2-(dx^2+dy^2+dz^2).\eqno(1.11)$$
The cosmological flat spacevelocity line element is given, accordingly, by
$$ds^2=\tau^2dv^2-(dx^2+dy^2+dz^2).\eqno(1.12)$$

The special relativistic line element is invariant under the Lorentz 
transformation. So is the cosmological line element: it is invariant under the
Lorentz-like cosmological transformation. The first keeps invariant the propagation 
of light, whereas the second keeps invariant the expansion of the universe.
At small velocities with respect to the speed of light, $v\ll c$, the Lorentz
transformation goes over to the nonrelativistic Galileo transformation. So is
the situation in cosmology: the Lorentz-like cosmological transformation goes over 
to the nonrelativistic Hubble transformation (see Subsection A) that is valid 
for cosmic times much smaller than the Hubble time, $t\ll\tau$.
\subsection{Five-Dimensional Manifold of Space, Time and Velocity}
If we add the time to the cosmological flat spacevelocity line element, we obtain
$$ds^2=c^2dt^2-(dx^2+dy^2+dz^2)+\tau^2 dv^2.\eqno(1.13)$$
Accordingly, we have a five-dimensional manifold of time, space and velocity.
The above line element provides a group of transformations O(2,3). At 
$v$=const it yields the Minkowskian line element (1.11); at $t$=const it gives
the cosmological line element (1.12); and at a fixed space point, $dx=dy=dz=0$,
it leads to a new two-dimensional line element
$$ds^2=c^2dt^2+\tau^2 dv^2.\eqno(1.14)$$
The groups associated with the above line elements are, of course, O(1,3),
O(3,1) and O(2), respectively. They are the Lorentz group, the cosmological
group and a two-dimensional Euclidean group, respectively.

In Section II we discuss some properties of the universe with gravitation in
five dimensions. That includes the Bianchi identities, the gravitational field
equations, the velocity as an independent coordinate and the enrgy density in
cosmology. In Section III we find the equations of motion of the expanding
universe and show that the universe is accelerating. In Section IV we discuss 
the important problem of halo dark matter around galaxies by finding the 
equations of motion of a star moving around a 
spherically-symmetric galaxy. The equations obtained are {\it not} Newtonian 
and instead the
Tully-Fisher formula is obtained from our theory. In Section V the cosmological 
constant is discussed. Our theory predicts that $\Lambda\approx 3H_0^2\approx
1.934\times 10^{-35}$s$^{-2}$, in agreement with the supernovae experiments 
teams. In Section VI once again we show that the 
universe is infinite and open, now by applying redshift analysis, using a 
new formula that is derived here. Section VII is devoted to  the concluding  
remarks.
\section{Universe with Gravitation}
The universe is, of course, not flat but filled up with gravity. When 
gravitation is invoked, the above spaces become curved Riemanian with the 
line element 
$$ds^2=g_{\mu\nu}dx^\mu dx^\nu, \eqno(2.1)$$
where $\mu$, $\nu$ take the values 0, 1, 2, 3, 4. The coordinates are: $x^0=
ct$, $x^1,x^2,x^3$ are spatial coordinates and $x^4=\tau v$ (the role of the
velocity as an independent coordinate will be discussed in the sequel). The 
signature is
$(+---+)$. The metric tensor $g_{\mu\nu}$ is symmetric and thus we have 
fifteen independent components. They will be a solution of the Einstein field 
equations in five dimensions. A discussion on the generalization of the 
Einstein field equations from four to five dimensions will also be given. 

The
five-dimensional field equations will not explicitely include a cosmological
constant. Our cosmological constant, extracted from the theory, will be equal 
to $\Lambda=3/\tau^2=1.934\times 
10^{-35}$s$^{-2}$ (for $H_0=73$km/s-Mpc). 
This should
be compared with results of the experiments recently done with the supernovae 
which suggest
the value of $\Lambda\approx 10^{-35}$s$^{-2}$. Our cosmological constant
is derived from the theory itself which is necessary to the classification of
the cosmological spaces to describe deccelerating, constant or accelerating
universe. We now discuss some basic questions that are encountered in going
from four to five dimensions.
\subsection{The Bianchi Identities} 
The restricted Bianchi identities are given by [2]
$$\left(R_\mu^\nu-\frac{1}{2}\delta_\mu^\nu R\right)_{;\nu}=0,\eqno(2.2)$$
where $\mu$,$\nu$=0,$\ldots$,4.
They are valid in five dimensions just as they are in four dimensions. In 
Eq. (2.2) $R_\mu^\nu$ and $R$ are the Ricci tensor and scalar, respectively,
and a semicolon denotes covariant differentiation. As a consequence we now 
have five coordinate conditions which permit us to determine five coordinates.
For example, one can choose $g_{00}=1$, $g_{0k}=0$, $g_{44}=1$, where $k$=1, 2,
3. These are the co-moving coordinates in five dimensions that keep the clocks
and the velocity-measuring instruments synchronized. We will not use these
coordinates in this paper. 
\subsection{The Gravitational Field Equations}
In four dimensions these are the Einstein field equations [3]:
$$R_{\mu\nu}-\frac{1}{2}g_{\mu\nu}R=\kappa T_{\mu\nu},\eqno(2.3)$$
or equivalently 
$$R_{\mu\nu}=\kappa\left(T_{\mu\nu}-\frac{1}{2}g_{\mu\nu}T\right),\eqno(2.4)$$
where $T=g_{\alpha\beta}T^{\alpha\beta}$, and we have $R=-\kappa T$. 
In five dimensions if one chooses Eq. (2.3) as the field equations then Eq. 
(2.4) is {\it not} valid (the
factor $\frac{1}{2}$ will have to be replaced by $\frac{1}{3}$, and $R=-
\frac{2}{3}\kappa T$), and thus there is no symmetry between $-\kappa T$ and $R$.

In fact, if one assumes Eq. (2.4) to be valid in five dimensions then a 
simple calculation shows that $\kappa T=-\frac{2}{3}R$ and Eq. (2.4) becomes
$$R_{\mu\nu}-\frac{1}{3}g_{\mu\nu}R=\kappa T_{\mu\nu}.\eqno(2.5)$$
Using now the Bianchi identities (2.2) then leads to $\partial R/\partial x^\nu
=0$ and thus $\partial T/\partial x^\nu=0$. If one chooses, for example, the
expression for the energy-momentum tensor to be given by $T^{\mu\nu}=\rho
\left(dx^\mu/ds\right)\left(dx^\nu/ds\right)$, then $T=T^{\mu\nu}g_{\mu\nu}=
\rho$. But if $T$ is a constant then $\rho$ is a constant (independent of 
time). This is obviously unacceptable situation for the universe since $\rho$
decreases as the universe expands. Hence Eq. (2.4) has to be rejected on
physical grounds, and the field equations that will be used by us in five
dimensions are those given by (2.3). 

Finally it is worthwhile mentioning that the only field equations in five
dimensions that have symmetry between geometry and matter like those in the
Einstein field equations in four dimensions are:
$$R_\mu^\nu-\frac{2}{5}\delta_\mu^\nu R=\kappa T_\mu^\nu,\eqno(2.6)$$
$$R_\mu^\nu=\kappa\left(T_\mu^\nu-\frac{2}{5}\delta_\mu^\nu T\right),
\eqno(2.7)$$
with $R=-\kappa T$.
While these equations are interesting, however, they do not reduce to Newtonian
gravity and thus they do not seem to be of physical interest.   
\subsection{Velocity as an Independent Coordinate}
First we have to iterate what do we mean by coordinates in general and how
one measures them. The time coordinate is measured by clocks as was emphasized
by Einstein repeatedly [4,5]. So are the spatial coordinates: they are 
measured by meters,
as was originally done in special relativity theory by Einstein, or by use of
Bondi's more modern version of k-calculus [6,7]. 

But how about the velocity as an
independent coordinate? One might incline to think that if we know the
spatial coordinates then the velocities are just the time derivative of the
coordinates and they are not independent coordinates. This is, indeed, the
situation for a dynamical system when the coordinates are given as functions 
of the time. But in general the situation is different, especially in 
cosmology. Take, for instance, the Hubble law $v=H_0x$. Obviously $v$ and $x$
are independent parameters and $v$ is not the time derivative of $x$. Basically
one can measure $v$ by instruments like those used by traffic police. 
\subsection{Effective Mass Density in Cosmology} 
To finish this section we discuss the important concept of the energy density
in cosmology. We use the Einstein field equations, in which the right-hand 
side includes the energy-momentum tensor. For fields other than gravitation,
like the electromagnetic field, this is a straightforward expression that
comes out as a generalization to curved spacetime of the same tensor 
appearing in special-relativistic electrodynamics. However, when dealing with
matter one should construct the energy-momentum tensor according to the 
physical situation (see, for example, Fock, Ref. 26). Often a special 
expression for the
mass density $\rho$ is taken for the right-hand side of Einstein's equations, 
which sometimes is expressed as a $\delta$-function. 

In cosmology we also have the situation where the mass density is put on the
right-hand side of the Einstein field equations. There is also the (constant) 
critical mass density $\rho_c=3/8\pi G\tau^2$, the value of which is about 
$10^{-29}$
g/cm$^3$, just a few hydrogen atoms per cubic meter throughout the cosmos. If
the universe average mass density $\rho$ is equal to $\rho_c$ then the three
spatial geometry of the four-dimensional cosmological space is Euclidian. A
deviation from this Euclidian geometry necessiates an increase or decrease of
$\rho_c$. That is to say that 
$$\rho_{eff}=\rho-\rho_c\eqno(2.8)$$ 
is the active or the effective 
mass density that causes the three geometry not to be Euclidian. Accordingly,
one should use $\rho_{eff}$ in the right-hand side of the Einstein  field 
equations. Indeed, we will use such a convention throughout this paper. The
subtraction of $\rho_c$ from $\rho$ in not significant for celestial bodies
amd makes no difference.  
\section{The Accelerating Universe}
\subsection{Preliminaries}
In the last two sections we gave arguments to the fact that the universe 
should be presented in five dimensions, even though the standard cosmological
theory is obtained from Einstein's four-dimensional general relativity theory.
The situation here is similar to that prevailed before the advent of ordinary
special relativity. At that time the equations of electrodynamics, written in
three dimensions, were well 
known to predict that the speed of light was constant. But that was not the 
end of the road. The abandon of the concept of absolute space along with the
constancy of the speed of light led to the four-dimensional notion. In cosmology
now, we have to give up the notion of absolute cosmic time. Then this with the
constancy of the Hubble constant in the limit of zero distance leads us to a
five-dimensional presentation of cosmology.  

We recall that the field equations are those of Einstein in five dimensions,
$$R_\mu^\nu-\frac{1}{2}\delta_\mu^\nu R=\kappa T_\mu^\nu, \eqno(3.1)$$
where Greek letters $\alpha,\beta,\cdots,\mu,\nu,\cdots=0,1,2,3,4$. The 
coordinates are $x^0=ct$, $x^1$, $x^2$ and $x^3$ are space-like coordinates,
$r^2=(x^1)^2+(x^2)^2+(x^3)^2$, and $x^4=\tau v$. The metric used is
given in an approximate form by (see Appendix A)
$$g_{\mu\nu}=\left(\begin{array}{ccccc}
1+\phi&0&0&0&0\\
0&-1&0&0&0\\
0&0&-1&0&0\\
0&0&0&-1&0\\
0&0&0&0&1+\psi\\\end{array}\right),\eqno(3.2)$$
We will keep only linear terms. The nonvanishing Christoffel symbols are given 
by (see Appendix A)
$$\Gamma^0_{0\lambda}=\frac{1}{2}\phi_{,\lambda},\hspace{1mm}
\Gamma^0_{44}=-\frac{1}{2}\psi_{,0}, \hspace{1mm}
\Gamma^n_{00}=\frac{1}{2}\phi_{,n}, \hspace{1mm}
\Gamma^n_{44}=\frac{1}{2}\psi_{,n},\hspace{1mm}
\Gamma^4_{00}=-\frac{1}{2}\phi_{,4},\hspace{1mm}
\Gamma^4_{4\lambda}=\frac{1}{2}\psi_{,\lambda},\eqno(3.3)$$
where $n=1,2,3$ and a comma denotes partial differentiation. The 
components of the Ricci tensor and the Ricci scalar are given by (Appendix B) 
$$R_0^0=\frac{1}{2}\left(\nabla^2\phi-\phi_{,44}-\psi_{,00}\right),
\eqno(3.4a)$$
$$R_0^n=\frac{1}{2}\psi_{,0n},\hspace{5mm} R_n^0=-\frac{1}{2}\psi_{,0n},
\hspace{5mm} R_0^4=R_4^0=0,\eqno(3.4b)$$
$$R_m^n=\frac{1}{2}\left(\phi_{,mn}+\psi_{,mn}\right),\eqno(3.4c)$$
$$R_n^4=-\frac{1}{2}\phi_{,n4},\hspace{5mm} R_4^n=\frac{1}{2}\phi_{,n4}.
\eqno(3.4d)$$
$$R_4^4=\frac{1}{2}\left(\nabla^2\psi-\phi_{,44}-\psi_{,00}\right),
\eqno(3.4e)$$
$$R=\nabla^2\phi+\nabla^2\psi-\phi_{,44}-\psi_{,00}.\eqno(3.5)$$
In the above equations $\nabla^2$ is the ordinary Laplace operator. 
\subsection{Expanding Universe}
The line element in five dimensions is given by 
$$ds^2=(1+\phi)dt^2-dr^2+(1+\psi)dv^2,\eqno(3.6)$$
where $dr^2=(dx^1)^2+(dx^2)^2+(dx^3)^2$, and where $c$ and $\tau$ were taken,
for brevity, as equal to 1. For an expanding universe one has $ds=0$. The 
line element (3.6) represents a spherically symmetric universe.

The expansion of the universe (the Hubble expansion) is recorded at a definite 
fixed time and thus $dt=0$. Accordingly Eq. (3.6) gives the following equation
for the (spherically symmetric) expansion of the universe at a certain moment,
$$-dr^2+(1+\psi)dv^2=0,\eqno(3.7)$$ 
and thus 
$$\left(\frac{dr}{dv}\right)^2=1+\psi.\eqno(3.8)$$
To find $\psi$ we solve the Einstein field equation (noting that $T_0^0=
g_{0\alpha}T^{\alpha 0}\approx T^{00}=\rho(dx^0/ds)^2\approx c^2\rho$, or 
$T^0_0\approx\rho$ in units with $c=1$):  
$$R_0^0-\frac{1}{2}\delta_0^0R=8\pi G\rho_{eff}=8\pi G\left(\rho-\rho_c\right),\eqno(3.9)$$ 
where $\rho_c=3/8\pi G\tau^2$. 

A simple calculation using Eqs. (3.4a) and (3.5) then yields 
$$\nabla^2\psi=6(1-\Omega_M), \eqno(3.10)$$
where $\Omega_M=\rho/\rho_c$.

The solution of the field equation (3.10) is given by 
$$\psi=(1-\Omega_M)r^2+\psi_0, \eqno(3.11)$$
where the first part on the right-hand side is a solution for the 
non-homogeneous Eq. (3.10), and $\psi_0$ represents a solution to its 
homogeneous part, i.e. $\nabla^2\psi_0=0$. A solution for $\psi_0$ can be
obtained as an infinite series in powers of $r$. The only term that is left is
of the form $\psi_0=-K_2/r$, where $K_2$ is a constant whose value can easily 
be shown to be the Schwartzschild radius, $K_2=2GM$. We therefore have
$$\psi=(1-\Omega_M)r^2-2GM/r.\eqno(3.12)$$  

The universe expansion is therefore given by 
$$\left(\frac{dr}{dv}\right)^2=1+\left(1-\Omega_M\right)r^2-\frac{2GM}{r}.
\eqno(3.13)$$   
For large $r$ the last term on the right-hand side of (3.13) can be neglected,
and therefore 
$$\left(\frac{dr}{dv}\right)^2=1+(1-\Omega_M)r^2,\eqno(3.14)$$
or 
$$\frac{dr}{dv}=\left[1+\left(1-\Omega_M\right)r^2\right]^{1/2}.\eqno(3.15)$$
Inserting now the constants $c$ and $\tau$ we finally obtain for the expansion
of the universe
$$\frac{dr}{dv}=\tau\left[1+\left(1-\Omega_M\right)r^2/c^2\tau^2\right]^{1/2}.
\eqno(3.16)$$
This result is exactly that obtained by Behar and Carmeli (BC) (Eq. 5.10) when 
the non-relativistic  relation $z=v/c$, where $z$ is the redshift parameter, 
is inserted in the previous result [8].

The second term in the square brackets of (3.16) represents the deviation
from constant expansion due to gravity. For without this term, Eq. (3.16)
reduces to $dr/dv=\tau$, thus $r=\tau v+$const. The constant can be taken zero
if one assumes, as usual, that at $r=0$ the velocity should also vanish. 
Accordingly we have $r=\tau v$ or $v=\tau^{-1}r$. Hence when 
$\Omega_M=1$, namely when $\rho=\rho_c$, we have a constant expansion.
\subsection{Decelerating, Constant and Accelerating Expansions}
The equation of motion (3.16) can be integrated exactly (see Appendix C). The
results are: \newline
For the $\Omega_M>1$ case 
$$r(v)=\left(c\tau/\alpha\right)\sin\left(\alpha v/c\right); \hspace{5mm}
\alpha=\left(\Omega_M-1\right)^{1/2}.\eqno(3.17)$$
This is obviously a decelerating expansion.\newline
For $\Omega_M<1$,
$$r(v)=\left(c\tau/\beta\right)\sinh\left(\beta v/c\right); \hspace{5mm}
\beta=\left(1-\Omega_M\right)^{1/2}.\eqno(3.18)$$
This is now an accelerating expansion. 

For $\Omega_M=1$ we have, from Eq. (3.16),
$$d^2r/dv^2=0,\eqno(3.19)$$
whose solution is, of course,
$$r(v)=\tau v,\eqno(3.20)$$
and this is a constant expansion. It will be noted that the last solution
can also be obtained directly from the previous two cases for $\Omega_M>1$ and
$\Omega_M<1$ by going to the limit $v\rightarrow 0$, using L'Hospital's lemma,
showing that our solutions are consistent. 

It has been shown in BC that the 
constant expansion is just a transition stage between the decelerating and the
accelerating expansions as the universe evolves toward its present situation.
That occured at about 5Gyr from the Big Bang at a time the cosmic radiation
temperature was 146K [8]. 
\subsection{Accelerating Universe}
In order to decide which of the three cases is the appropriate one at the present
time, it will be convenient to write the solutions (3.17), (3.18) and (3.20) in the ordinary
Hubble law form $v=H_0r$. Expanding Eqs. (3.17) and (3.18) and keeping the
appropriate terms then yields 
$$r=\tau v\left(1-\alpha^2v^2/6c^2\right),\eqno(3.21)$$
$$r=\tau v\left(1+\beta^2v^2/6c^2\right),\eqno(3.22)$$ 
for the $\Omega_M>1$ and $\Omega_M<1$ cases, respectively.
Using now the expressions for $\alpha$ and $\beta$ in Eqs. (3.21) and (3.22),
then both of the latter can be reduced into the single equation
$$r=\tau v\left[1+\left(1-\Omega_M\right)v^2/6c^2\right].\eqno(3.23)$$
Inverting now this equation by writing it in the form $v=H_0r$, we obtain in
the lowest approximation for $H_0$ 
$$H_0=h\left[1-\left(1-\Omega_M\right)v^2/6c^2\right],\eqno(3.24)$$
where $h=1/\tau$. Using $v\approx r/\tau$, or $z\approx v/c$, we also obtain
$$H_0=h\left[1-\left(1-\Omega_M\right)r^2/6c^2\tau^2\right]=h\left[1-\left(1-
\Omega_M\right)z^2/6\right].\eqno(3.25)$$
 
As is seen $H_0$ depends on the distance, or equivalently, on the redshift.
Cosequently $H_0$ has meaning only in the limits $r\rightarrow 0$ and $z
\rightarrow 0$, namely when measured {\it locally}, in which case it becomes 
the constant $h$. This is similar to the situation with respect to the speed 
of light when measured globally in the presence of gravitational field as the 
ratio between distance and time, the result usually depends on these 
parameters. Only in the limit one obtains the constant speed of light in 
vacuum ($c\approx 3\times 10^{10}$cm/s). 

As is seen from the above discussion, $H_0$ is intimately related to the sign
of the factor $(1-\Omega_M)$. If measurements of $H_0$ indicate that it 
increases with the redshift parameter $z$ then the sign of $(1-\Omega_M)$ is 
negative, namely $\Omega_M>1$. If, however, $H_0$ decreases when $z$ increases
then the sign of $(1-\Omega_M)$ is positive, i.e. $\Omega_M<1$. The possibility of
$H_0$ not to depend on the redshift parameter indicates that $\Omega_M=1$. In
recent years different measurements were obtained for $H_0$, with the 
so-called 
``short" and ``long" distance scales, in which higher values of $H_0$ were 
obtained for the short distances and the lower values for $H_0$ corresponded 
to the long distances [9-18]. Indications are that the longer the distance of 
measurement,
the smaller the value of $H_0$. If one takes these experimental results 
seriously, then that is possible only for the case in
which $\Omega_M<1$, namely when the universe is at an accelerating expansion 
phase, and the universe is thus open. We will see in Section VI that
the same result is obtained via a new cosmological redshift formula.  
\section{The Tully-Fisher Formula: Nonexistence of Halo Dark Matter?}
In this section we derive the equations of motion of a star moving around a 
spherically symmetric galaxy and show that the Tully-Fisher formula is 
obtained from our five-dimensional cosmological theory. The calculation is
lengthy but it is straightforward. The equations of 
motion
will first be of general nature and only afterward specialized to the motion 
of a star around the field of a galaxy. The equations obtained are {\it not} 
Newtonian. The Tully-Fisher formula was obtained by us in a previous paper [19]
using two representations of Einstein's general relativity: the 
standard spacetime theory and a spacevelocity version of it. However, the
present derivation is a straightforward result from the unification of space, 
time and velocity.   

Our notation in this section is as follows: $\alpha,\beta,\gamma,\dots=0,\cdots,
4$; $a,b,c,d,\cdots=0,\cdots,3$; $p,q,r,s,\cdots=1,\cdots,4$; and $k,l,m,n,
\cdots=1,2,3$. The coordinates are: $x^0=ct$ (timelike), $x^k=x^1,x^2,x^3$ 
(spacelike), and $x^4=\tau v$ (velocitylike -- see Section 2C). 
\subsection{The Geodesic Equation}
As usual the equations of motion are obtained in general relativity theory
from the covariant conservation law of the energy-momentum tensor (which is
a consequence of the restricted Bianchi identities), and the 
result, as is well known, is the geodesic equation that describes the motion 
of a spherically symmetric test particle. In our five-dimensional cosmological
model we have five equations of motion. They are given by 
$$\frac{d^2x^\mu}{ds^2}+\Gamma_{\alpha\beta}^\mu\frac{dx^\alpha}{ds}
\frac{dx^\beta}{ds}=0.\eqno(4.1)$$

We now change the independent parameter $s$ into an arbitrary new parameter 
$\sigma$, then the geodesic equation becomes
$$\frac{d^2x^\mu}{d\sigma^2}+\Gamma_{\alpha\beta}^\mu\frac{dx^\alpha}{d\sigma}
\frac{dx^\beta}{d\sigma}=-\frac{d^2\sigma/ds^2}{\left(d\sigma/ds\right)^2}
\frac{dx^\mu}{d\sigma}.\eqno(4.2)$$
The parameter $\sigma$ will be taken once as $\sigma=x^0$ (the 
time coordinate) and then $\sigma=x^4$ (the velocity coordinate). We obtain,
for the first case,
$$\frac{d^2x^\mu}{\left(dx^0\right)^2}+\Gamma_{\alpha\beta}^\mu\frac{dx^\alpha}{dx^0}
\frac{dx^\beta}{dx^0}=-\frac{d^2x^0/ds^2}{\left(dx^0/ds\right)^2}    
\frac{dx^\mu}{dx^0}.\eqno(4.3)$$  
The right-hand side of Eq. (4.3) can be written in a somewhat different form
by using its zero component
$$\frac{d^2x^0}{\left(dx^0\right)^2}+\Gamma_{\alpha\beta}^0\frac{dx^\alpha}{dx^0}
\frac{dx^\beta}{dx^0}=-\frac{d^2x^0/ds^2}{\left(dx^0/ds\right)^2}    
\frac{dx^0}{dx^0}.\eqno(4.4)$$
But $dx^0/dx^0=1$, and $d^2x^0/(dx^0)^2=0$. Hence we obtain
$$\frac{d^2x^0/ds^2}{\left(dx^0/ds\right)^2}=-\Gamma_{\alpha\beta}^0
\frac{dx^\alpha}{dx^0}\frac{dx^\beta}{dx^0}.\eqno(4.5)$$

Using the above result in Eq. (4.3), the latter can be writtem in the form
$$\frac{d^2x^\mu}{\left(dx^0\right)^2}+\left(\Gamma^\mu_{\alpha\beta}-
\Gamma^0_{\alpha\beta}\frac{dx^\mu}{dx^0}\right)\frac{dx^\alpha}{dx^0}
\frac{dx^\beta}{dx^0}=0.\eqno(4.6)$$
It will be noted that the zero component Eq. (4.6) is now an identity, and
consequently it reduces to the four-dimensional equation
$$\frac{d^2x^p}{\left(dx^0\right)^2}+\left(\Gamma^p_{\alpha\beta}-
\Gamma^0_{\alpha\beta}\frac{dx^p}{dx^0}\right)\frac{dx^\alpha}{dx^0}
\frac{dx^\beta}{dx^0}=0,\eqno(4.7)$$
where $p=1,2,3,4$.

In exactly the same way we parametrize the geodesic equation (4.1) now with
respect to the velocity by choosing the parameter $\sigma=\tau v$. The result
is 
$$\frac{d^2x^a}{\left(dx^4\right)^2}+\left(\Gamma^a_{\alpha\beta}-            
\Gamma^4_{\alpha\beta}\frac{dx^a}{dx^4}\right)\frac{dx^\alpha}{dx^4}         
\frac{dx^\beta}{dx^4}=0,\eqno(4.8)$$ 
where $a=0,1,2,3$.

The equation of motion (4.7) will be expanded in terms of the parameter $v/c$,
assuming $v\ll c$, whereas Eq. (4.8) will be expanded with $t/\tau$, where
$t$ is a characteristic cosmic time, and $t\ll\tau$. We then can use the
Einstein-Infeld-Hoffmann (EIH) method that is well known in general relativity
in obtaining the equations of motion [20-37]. 

We start with Eq. (4.7). As is seen the
second term in the paranthesis can be neglected with respect to the first one
since $d/dx^0=(1/c)d/dt$, and we obtain    
$$\frac{d^2x^p}{\left(dx^0\right)^2}+\Gamma^p_{\alpha\beta}
\frac{dx^\alpha}{dx^0}\frac{dx^\beta}{dx^0}=0.\eqno(4.9)$$
In Eq. (4.8) we also can neglect the second term in the paranthesis since
$d/dx^4=(1/\tau)d/dv$. As a result we have the approximate equations of
motion
$$\frac{d^2x^a}{\left(dx^4\right)^2}+\Gamma^a_{\alpha\beta}
\frac{dx^\alpha}{dx^4}\frac{dx^\beta}{dx^4}=0.\eqno(4.10)$$ 
The equations of motion (4.9) and (4.10) are consequently given by
$$\frac{d^2x^p}{dt^2}+\Gamma^p_{\alpha\beta}
\frac{dx^\alpha}{dt}\frac{dx^\beta}{dt}=0,\eqno(4.11)$$
$$\frac{d^2x^a}{dv^2}+\Gamma^a_{\alpha\beta}
\frac{dx^\alpha}{dv}\frac{dx^\beta}{dv}=0.\eqno(4.12)$$

To find the lowest approximation of Eq. (4.11), since $dx^0/dt\gg 
dx^q/dt$, all terms with indices that are not zero-zero can be neglected. 
Consequently, Eq. (4.11) is reduced to the form                             
$$\frac{d^2x^p}{dt^2}\approx -\Gamma^p_{00},\eqno(4.13)$$
in the lowest approximation. 
\subsection{Equations of Motion}
Accordingly $\Gamma^p_{00}$ acts like a Newtonian force per mass unit. In 
terms of the metric tensor we therefore obtain, since $\Gamma^p_{00}=
-\frac{1}{2}\eta^{pq}\phi_{,q}$ (see Appendix A)
$$\frac{d^2x^p}{dt^2}\approx -\frac{1}{2}\eta^{pq}
\frac{\partial\phi}{\partial q},\eqno(4.14)$$
where $\phi=g_{00}-1$ (see Appendix A). We now decompose this equation into
a spatial ($p=1,2,3$) and a velocity ($p=4$) parts, getting 
$$\frac{d^2x^k}{dt^2}=-\frac{1}{2}\frac{\partial\phi}{\partial x^k},
\eqno(4.15a)$$
$$\frac{d^2v}{dt^2}=0.\eqno(4.15b)$$
Using exactly the same method, Eq. (4.12) yields 
$$\frac{d^2x^k}{dv^2}=-\frac{1}{2}\frac{\partial\psi}{\partial x^k},
\eqno(4.16a)$$ 
$$\frac{d^2t}{dv^2}=0,\eqno(4.16b)$$
where $\psi=g_{44}-1$. 
In the above equations $k=1,2,3$. Equation (4.15a) is exactly the law of 
motion with the function $\phi$ being twice the Newtonian potential. The other
three equations Eq. (4.15b) and Eqs. (4.16a,b) are not Newtonian and are
obtained only in the present theory. It remains to find out the functions
$\phi$ and $\psi$. 

To find out the function $\phi$ we solve the Einstein field equation (noting
that $T_4^4=g_{4\alpha}T^{\alpha 4}\approx T^{44}=\rho(dx^4/ds)^2\approx\tau^2
\rho$, and thus $T_4^4\approx\rho$ in units in which $\tau=1$):
$$R_4^4-\frac{1}{2}\delta_4^4 R=8\pi G\rho_{eff}=8\pi G(\rho-\rho_c).\eqno(4.17)$$
A straightforward calculation then gives 
$$\nabla^2\phi=6\left(1-\Omega_M\right),\eqno(4.18)$$
whose solution is given by
$$\phi=\left(1-\Omega_M\right)r^2+\phi_0,\eqno(4.19)$$
where $\phi_0$ is a solution of the homogeneous equation $\nabla^2\phi_0=0$.
One then easily finds that $\phi_0=-K_1/r$, where $K_1=2GM$. In the same way
the function $\psi$ can be found (see Section 3),
$$\psi=\left(1-\Omega_M\right)r^2+\psi_0,\eqno(4.20)$$
with $\nabla^2\psi_0=0$, $\psi_0=-K_2/r$ and $K_2=2GM$. (When units are
inserted then $K_1=2GM/c^2$ and $K_2=2GM\tau^2/c^2$.) For the purpose of 
obtaining equations of motion one can neglect the terms $(1-\Omega_M)r^2$, 
actually $(1-\Omega_M)r^2/c^2\tau^2$, in the solutions for $\phi$ and $\psi$.
One then obtains 
$$g_{00}\approx 1-2GM/c^2r,\hspace{5mm}g_{44}\approx 1-2GM\tau^2/c^2r.\eqno(4.21)$$

The equations of motion (4.15a) and (4.16a), consequently, have the forms
$$\frac{d^2x^k}{dt^2}=\left(\frac{GM}{r}\right)_{,k},\hspace{5mm}
\frac{d^2x^k}{dv^2}=\left(\frac{GM}{r}\right)_{,k},\eqno(4.22)$$
or, when inserting the constants $c$ and $\tau$, 
$$\frac{d^2x^k}{dt^2}=GM\left(\frac{1}{r}\right)_{,k},\eqno(4.23a)$$
$$\frac{d^2x^k}{dv^2}=kM\left(\frac{1}{r}\right)_{,k},
\eqno(4.23b)$$
where $k=G\tau^2/c^2$.
It remains to integrate equations (4.15b) and (4.16b). One finds that $v=a_0t$,
where $a_0$ is a constant which can be taken as equal to $a_0=c/\tau\approx
cH_0$. 
Accordingly, we see that the particle experiences an acceleration $a_0=c/\tau
\approx cH_0$ directed outward when the motion is circular.

Equation (4.23a) is Newtonian but (4.23b) is not. The integration of the
latter is identical to that familiar in classical Newtonian mechanics, but
there is an essential difference which should be emphasized. 
In Newtonian equations of motion one deals with a path of motion in the 
3-space. In our theory we do not have that situation. Rather, the paths here
indicate locations of particles in the sense of the Hubble distribution, which
now takes a different physical meaning. With that in mind we proceed as
follows. 

Equation (4.23b) yields the first integral
$$\left(\frac{ds}{dv}\right)^2=\frac{kM}{r},\eqno(4.24a)$$
where $v$ is the velocity of the particles, in analogy to the Newtonian
$$\left(\frac{ds}{dt}\right)^2=\frac{GM}{r},\eqno(4.24b)$$
In these equations $s$ is the length parameter along the path of the accumulation
of the particles.

Comparing Eqs. (4.24a) and (4.24b), we obtain
$$\frac{ds}{dv}=\frac{\tau}{c}\frac{ds}{dt}.\eqno(4.25)$$
Thus
$$\frac{dv}{dt}=\frac{c}{\tau}.\eqno(4.26)$$
Accordingly, as we have mentioned before, the particle experiences 
an acceleration $a_0=c/\tau\approx
cH_0$ directed outward when the motion is circular.
\subsection{The Tully-Fisher Law}
The motion of a particle in a central field is best described in terms of an
``effective potential", $V_{eff}$. In Newtonian mechanics this is given by [38]
$$V_{eff}=-\frac{GM}{r}+\frac{L^2}{2r^2},\eqno(4.27)$$
where $L$ is the angular momentum per mass unit. In our case the effective
potential is
$$V_{eff}\left(r\right)=-\frac{GM}{r}+\frac{L^2}{2r^2}+a_0r.\eqno(4.28)$$

The circular motion is obtained at the minimal value of (4.28), i.e.,
$$\frac{dV_{eff}}{dr}=\frac{GM}{r^2}-\frac{L^2}{r^3}+a_0=0,\eqno(4.29)$$
with $L=v_cr$, and $v_c$ is the circular velocity. This gives
$$v_c^2=\frac{GM}{r}+a_0r.\eqno(4.30)$$
Thus
$$v_c^4=\left(\frac{GM}{r}\right)^2+2GMa_0+a_0^2r^2,\eqno(4.31)$$
where $a_0=c/\tau\approx cH_0$.

The first term on the right-hand side of Eq. (4.31) is purely Newtonian, and
cannot be avoided by any reasonable theory. The second one is the Tully-Fisher
term. The third term is extremely small at the range of distances of stars
around a galaxy. It is well known that astronomical observations show that
for disk galaxies the fourth power of the circular velocity of stars moving
around the core of the galaxy, $v_c^4$, is proportional to the total 
luminosity $L$ of the galaxy to an occuracy of more than two orders of 
magnitude in $L$, namely $v_c^4\propto L$ [39]. Since $L$ is proportional to the
mass $M$ of the galaxy, one obtains $v_c^4\propto M$. This is the Tully-Fisher
law. There is no dependence on the distance of the star from the center of the
galaxy as Newton's law $v_c^2=GM/r$ requires for circular motion. In order to
rectify this deviation from Newton's laws, astronomers assume the existence of
halos around the galaxy which are filled with dark matter and arranged in such
a way so as to satisfy the Tully-Fisher law for each particular situation. 

In conclusion it appears that there is no necessity for the assumption of the 
existence of halo 
dark matter around galaxies. Rather, the result can be described in terms of
the properties of spacetimevelocity.
\section{The Cosmological Constant}
\subsection{The Cosmological Term}
First, a historical remark. In order to allow the existence of a static
solution for  the gravitational field equations, Einstein made a modification
to his original equations (2.3) by adding a cosmological term, 
$$R_{\mu\nu}-\frac{1}{2}g_{\mu\nu}R+\Lambda g_{\mu\nu}=\kappa T_{\mu\nu},
\eqno(5.1)$$
where $\Lambda$ is the cosmological constant and $\kappa=8\pi G$. For a homogeneous and 
isotropic universe with the line element [40,41]
$$ds^2=dt^2-a^2\left(t\right)R_0^2\left[\frac{dr^2}{1-kr^2}+r^2\left(
d\theta^2+\sin^2\theta d\phi^2\right)\right],\eqno(5.2)$$
where $k$ is the curvature parameter ($k=1,0,-1$) and $a(t)=R(t)/R_0$ is the 
scale factor, with the energy-momentum tensor 
$$T_{\mu\nu}=\left(\rho+p\right)u_\mu u_\nu-pg_{\mu\nu},\eqno(5.3)$$
Einstein's equations (5.1) reduce to the two Friedmann equations
$$H^2\equiv\left(\frac{\dot{a}}{a}\right)^2=\frac{\kappa}{3}\rho+
\frac{\Lambda}{3}-\frac{k}{a^2R_0^2},\eqno(5.4)$$
$$\frac{\ddot{a}}{a}=-\frac{\kappa}{6}\left(\rho+3p\right)+\frac{\Lambda}{3}.
\eqno(5.5)$$
These equations admit a static solution $(\dot{a}=0)$ with $k>0$ and 
$\Lambda>0$. After Hubble's discovery that the universe is expanding, the role
of the cosmological constant to allow static homogeneous solutions to Einstein's
equations in the presence of matter, seemed to be unnecessary. For a long
time the cosmological term was considered to be of no physical interest in 
cosmological problems.

From the Friedmann equation (5.4), for any value of the Hubble parameter $H$
there is a critical value of the mass density such that the spatial geometry
is flat ($k=0$), $\rho_c=3H_0^2/\kappa$ (see Subsection IID). One usually 
measures the total
mass density in terms of the critical density $\rho_c$ by means of the 
density parameter $\Omega_M=\rho/\rho_c$. 

In general, the mass density $\rho$ includes contributions from various
distinct components. From the point of view of cosmology, the relevant aspect 
of each component is how its energy density evolves as the universe expands.
In general, a positive $\Lambda$ causes acceleration to the universe expansion,
whereas a negative $\Lambda$ and ordinary matter tend to decelerate it. 
Moreover, the relative contributions of the components to the energy density
change with time. For $\Omega_\Lambda<0$, the universe will always recollapse
to a Big Crunch. For $\Omega_\Lambda>0$ the universe will expand forever
unless there is sufficient matter to cause recollapse before $\Omega_\Lambda$
becomes dynamically important. For $\Omega_\Lambda=0$ we have the familiar
situation in which $\Omega_M\leq 1$ universes expand forever and $\Omega_M>1$
universes recollapse. (For more details see the paper by Behar and Carmeli,
Ref. 8.)
\subsection{The Supernovae Experiments Value for $\Lambda$}
Recently two groups, the {\it Supernova Cosmology Project Collaboration} and 
the {\it High-Z Supernova Team Collaboration}, presented evidence that the 
expansion of the universe is accelerating [42-48].
These teams have measured the distances to cosmological supernovae by using 
the fact that the intrinsic luminosity of Type Ia supernovae is closely 
correlated with their decline rate from maximum brightness, which can be
independently measured. These measurements, combined with redshift data for
the supernovae, led to the prediction of an accelerating universe. Both
teams obtained 
$$\Omega_M\approx 0.3,\hspace{5mm}\Omega_\Lambda\approx 0.7,\eqno(5.6)$$
and strongly ruled out the traditional ($\Omega_M$, $\Omega_\Lambda$)=(1, 0)
universe. This value of the density parameter $\Omega_\Lambda$ corresponds to
a cosmological constant that is small but nonzero and positive,
$$\Lambda\approx 10^{-52}\mbox{\rm m}^{-2}\approx 10^{-35}\mbox{\rm s}^{-2}.
\eqno(5.7)$$
\subsection{The Behar-Carmeli Predicted Value for $\Lambda$}
In the paper of Behar and Carmeli a four-dimensional cosmological relativity 
theory that unifies space and velocity was proposed that
predicts the acceleration of the universe and hence it is equivalent to having
a positive value for $\Lambda$ in it. As is well known, in the traditional 
work of Friedmann when added to it a cosmological constant, the field 
equations obtained are highly complicated and no solutions have been obtained 
so far. Behar-Carmeli's theory, on the other hand, yields exact solutions and 
describes the universe as having a three-phase
evolution with a decelerating expansion followed by a constant and an
accelerating expansion, and it predicts that the universe is now in the 
latter phase. In the framework of this theory the zero-zero component of
Einstein's equations is written as
$$R^0_0-\frac{1}{2}\delta_0^0R=\kappa\rho_{eff}=\kappa\left(\rho-\rho_c\right),
\eqno(5.8)$$
where $\rho_c=3/\kappa\tau^2\approx 3H^2/\kappa$ is the critical mass density. 
Comparing Eq. (5.8)
with the zero-zero component of Eq. (5.1), one obtains the expression for the
cosmological constant in the Behar-Carmeli theory, 
$$\Lambda=\kappa\rho_c=3/\tau^2\approx 3H^2.\eqno(5.9)$$
\subsection{Comparison with Experiment}
To find out the numerical value of $\tau$ we use the relationship between
$h=\tau^{-1}$ and $H_0$ given by Eq. (3.25) (CR denote values according to
Cosmological Relativity): 
$$H_0=h\left[1-\left(1-\Omega_m^{CR}\right)z^2/6\right],\eqno(5.10)$$
where $z=v/c$ is the redshift and $\Omega_m^{CR}=\rho_M/\rho_c$ with 
$\rho_c=3h^2/8\pi G$. (Notice that our $\rho_c=1.194\times 10^{-29}g/cm^3$ is different from 
the standard $\rho_c$ defined with $H_0$.) The redshift parameter $z$ 
determines the distance at which $H_0$ is measured. We choose $z=1$ and take 
for 
$$\Omega_m^{CR}=0.245, \eqno(5.11)$$
its value at the present time (corresponds to 0.32 in the 
standard theory), Eq. (5.10) then gives
$$H_0=0.874h.\eqno(5.12)$$
At the value $z=1$ the corresponding Hubble parameter $H_0$ according to the 
latest results from HST can be taken [9] as $H_0=70$km/s-Mpc, thus 
$h=(70/0.874)$km/s-Mpc, or
$$h=80.092\mbox{\rm km/s-Mpc},\eqno(5.13)$$
and 
$$\tau=12.486 Gyr=3.938\times 10^{17}s.\eqno(5.14)$$

What is left is to find the value of $\Omega_\Lambda^{CR}$. We have 
$\Omega_\Lambda^{CR}=\rho_c^{ST}/\rho_c$, where $\rho_c^{ST}=3H_0^2/8\pi 
G$ and $\rho_c=3h^2/8\pi G$. Thus $\Omega_\Lambda^{CR}=(H_0/h)^2=0.874^2$,
or
$$\Omega_\Lambda^{CR}=0.764.\eqno(5.15)$$
As is seen from Eqs. (5.11) and (5.15) one has 
$$\Omega_T=\Omega_m^{CR}+\Omega_\Lambda^{CR}=0.245+0.764=1.009\approx 1,
\eqno(5.16)$$
which means the universe is Euclidean.

As a final result we calculate the cosmological constant according to 
Eq. (5.9). One obtains
$$\Lambda=3/\tau^2=1.934\times 10^{-35}s^{-2}.\eqno(5.17)$$

Our results confirm those of the supernovae experiments and indicate on the
existance of the dark energy as has recently received confirmation from the
Boomerang cosmic microwave background experiment [50,51], which showed that 
the universe is Euclidean.
\section{Cosmological Redshift Analysis}
\subsection{The Redshift Formula}
In this section we derive a general formula for the redshift in which the
term $(1-\Omega_M)$ appears explicitly. Since there are enough data of 
measurements of redshifts, this allows one to determine what is the sign of
$(1-\Omega_M)$, positive, zero or negative. Our conclusion is that 
$(1-\Omega_M)$
cannot be negative or zero. This means that the universe is infinite, 
and expands forever, a result favored by some cosmologists [49]. To this end 
we proceed as follows.

Having the metric tensor from Section IV we may now find the redshift of light
emitted in the cosmos. As usual, at two points 1 and 2 we have for the wave 
lengths and frequencies:
$$\frac{\lambda_2}{\lambda_1}=\frac{\nu_1}{\nu_2}=
\frac{ds\left(2\right)}{ds\left(1\right)}=
\sqrt{\frac{g_{00}\left(2\right)}{g_{00}\left(1\right)}}.\eqno(6.1)$$ 
Using now the solution for $g_{00}=1+\phi$, with $\phi$ given by Eq. (4.19), 
in Eq. (6.1), we obtain 
$$\frac{\lambda_2}{\lambda_1}=
\sqrt{\frac{1+r_2^2/a^2-R_s/r_2}{1+r_1^2/a^2-R_s/r_1}}.\eqno(6.2)$$
In Eq. (6.2) $R_s=2GM/c^2$ and $a^2=c^2\tau^2/(1-\Omega_M)$.

For a sun-like body with radius $R$ located at the coordinates origin, and an
observer at a distance $r$ from the center of the body, we then have $r_2=r$
and $r_1=R$, thus
$$\frac{\lambda_2}{\lambda_1}=
\sqrt{\frac{1+r^2/a^2-R_s/r}{1+R^2/a^2-R_s/R}}.\eqno(6.3)$$
\subsection{Particular Cases}
Since $R\ll r$ and $R_s<R$ is usually the case we can write, to a good approximation,
$$\frac{\lambda_2}{\lambda_1}=
\sqrt{\frac{1+r^2/a^2}{1-R_s/R}}.\eqno(6.4)$$

The term $r^2/a^2$ in Eq. (6.4) is a pure cosmological one, whereas $R_s/R$
is the standard general relativistic term. For $R\gg R_s$ we then have
$$\frac{\lambda_2}{\lambda_1}=\sqrt{1+\frac{r^2}{a^2}}=
\sqrt{1+\frac{\left(1-\Omega_M\right)r^2}{c^2\tau^2}}\eqno(6.5)$$
for the pure cosmological contribution to the redshift. If, furthermore,
$r\ll a$ we then have
$$\frac{\lambda_2}{\lambda_1}=1+\frac{r^2}{2a^2}=
1+\frac{\left(1-\Omega_M\right)r^2}{2c^2\tau^2}\eqno(6.6)$$
to the lowest approximation in $r^2/a^2$, and thus
$$z=\frac{\lambda_2}{\lambda_1}-1=\frac{r^2}{2a^2}=
\frac{\left(1-\Omega_M\right)r^2}{2c^2\tau^2}.\eqno(6.7)$$

When the contribution of the cosmological term $r^2/a^2$ is negligible, we 
then have
$$\frac{\lambda_2}{\lambda_1}=\frac{1}{\sqrt{1-R_s/R}}.\eqno(6.8)$$
The redshift could then be very large if $R$, the radius of the emitting body,
is just a bit larger than the Schwarzchild radius $R_s$. For example if $R_s/R
=0.96$ the redshift $z=4$. For a typical sun like ours, $R_s\ll R$ and we can 
expand the righthand side of Eq. (6.8), getting
$$\frac{\lambda_2}{\lambda_1}=1+\frac{R_s}{2R},\eqno(6.9)$$
thus
$$z=\frac{R_s}{2R}=\frac{Gm}{c^2R},\eqno(6.10)$$
the standard general relativistic result.

From Eqs. (6.5)--(6.7) it is clear that $\Omega_M$ cannot be larger than one 
since otherwise $z$ will be negative, which means blueshift, and as is well
known nobody sees such a thing. If $\Omega_M=1$ then $z=0$, and for 
$\Omega_M<1$ we have $z>0$. The case of $\Omega_M=1$ is also implausible 
since the light from
stars we see is usually redshifted more than the redshift due to the gravity of the body
emitting the radiation, as is evident from our sun, for example, whose emitted
light is shifted by only $z=2.12\times 10^{-16}$ [3].
\subsection{Conclusions}
One can thus conclude that the theory of unified space, time and velocity 
predicts that the universe is open. As is well known the standard FRW model does not relate the 
cosmological redshift to the kind of universe.
\section{Concluding Remarks}
The most direct evidence that the universe expansion is accelerating and 
propelled by ``dark energy", is provided by the faintness of Type Ia supernovae 
(SNe Ia) at $z\approx 0.5$ [42,46]. Beyond the redshift range of $0.5<z<1$, the 
universe was more compact and the attraction of matter was stronger than the
repulsive dark energy. At $z>1$ the expansion of the universe should have been
decelerating [52]. At $z\geq 1$ one would expect an apparent brightness 
increase of SNe Ia relative to what is supposed to be for a non-decelerating 
universe [53].

Recently, more confirmation to the universe accelerating expansion came from 
the most distant supernova, SN 1997ff, that was recorded by the Hubble Space
Telescope. As has been pointed out before, if we look back far enough, we 
should find a decelerating expansion. Beyond $z=1$
one should see an earlier time when the mass density was dominant. The 
measurements obtained from SN 1997ff's redshift and brightness provide a 
direct proof for the transition from past decelerating to present 
accelerating expansion. The measurements also exclude the possibility that
the acceleration of the universe is not real but is due to other astrophysical 
effects such as dust. 
\section*{Appendix A: Mathematical Conventions and Christoffel Symbols}
Throughout this paper we use the convention 
$$\alpha,\beta,\gamma,\delta,\cdots=0,1,2,3,4,$$ 
$$a,b,c,d,\cdots=0,1,2,3,$$
$$p,q,r,s,\cdots=1,2,3,4,$$
$$k,l,m,n,\cdots=1,2,3.$$
The coordinates are $x^0=ct$, $x^1,x^2$ and $x^3$ are space-like coordinates,
$r^2=(x^1)^2+(x^2)^2+(x^3)^2$, and $x^4=\tau v$. The signature is $(+---+)$.
The metric, approximated up 
to $\phi$ and $\psi$, is:
$$g_{\mu\nu}=\left(\begin{array}{ccccc}
1+\phi&0&0&0&0\\
0&-1&0&0&0\\
0&0&-1&0&0\\
0&0&0&-1&0\\
0&0&0&0&1+\psi\\\end{array}\right)
,\eqno(A.1)$$
$$g^{\mu\nu}=\left(\begin{array}{ccccc}
1-\phi&0&0&0&0\\
0&-1&0&0&0\\
0&0&-1&0&0\\
0&0&0&-1&0\\
0&0&0&0&1-\psi\\\end{array}\right).
\eqno(A.2)$$
The nonvanishing Christoffel symbols are (in the linear approximation):
$$\Gamma^0_{0\lambda}=\frac{1}{2}\phi_{,\lambda},\hspace{5mm}
\Gamma^0_{44}=-\frac{1}{2}\psi_{,0},\hspace{5mm}
\Gamma^n_{00}=\frac{1}{2}\phi_{,n}, \eqno(A.3a)$$
$$\Gamma^n_{44}=\frac{1}{2}\psi_{,n},\hspace{5mm}
\Gamma^4_{00}=-\frac{1}{2}\phi_{,4},\hspace{5mm}
\Gamma^4_{4\lambda}=\frac{1}{2}\psi_{,\lambda},\eqno(A.3b)$$
$$\Gamma^a_{00}=-\frac{1}{2}\eta^{ab}\phi_{,b},\hspace{5mm}
\Gamma^a_{44}=-\frac{1}{2}\eta^{ab}\psi_{,b}, \eqno(A.3c)$$
$$\Gamma^p_{00}=-\frac{1}{2}\eta^{pq}\phi_{,q},\hspace{5mm}
\Gamma^p_{44}=-\frac{1}{2}\eta^{pq}\psi_{,q}. \eqno(A.3d)$$
The Minkowskian metric $\eta$ in five dimensions is given by
$$\eta=\left(\begin{array}{ccccc}
1&0&0&0&0\\
0&-1&0&0&0\\
0&0&-1&0&0\\
0&0&0&-1&0\\
0&0&0&0&1\\\end{array}\right).\eqno(A.4)$$
\section*{Appendix B: Components of the Ricci Tensor}
The elements of the Ricci tensor are:
$$R_{00}=\frac{1}{2}\left(\nabla^2\phi-\phi_{,44}-\psi_{,00}\right),\eqno(B.1)$$
$$R_{0n}=-\frac{1}{2}\psi_{,0n},\hspace{5mm}R_{04}=0,\eqno(B.2)$$
$$R_{mn}=-\frac{1}{2}\left(\phi_{,mn}+\psi_{,mn}\right),\eqno(B.3)$$
$$R_{4n}=-\frac{1}{2}\phi_{,4n},\eqno(B.4)$$
$$R_{44}=\frac{1}{2}\left(\nabla^2\psi-\phi_{,44}-\psi_{,00}\right).\eqno(B.5)$$
The Ricci scalar is
$$R=\nabla^2\phi+\nabla^2\psi-\phi_{,44}-\psi_{,00}.\eqno(B.6)$$
The mixed Ricci tensor is given by
$$R_0^0=\frac{1}{2}\left(\nabla^2\phi-\phi_{,44}-\psi_{,00}\right),
\eqno(B.7)$$
$$R_0^n=\frac{1}{2}\psi_{,0n},\hspace{5mm} R_n^0=-\frac{1}{2}\psi_{,0n},
\eqno(B.8)$$
$$R_0^4=R_4^0=0,\eqno(B.9)$$
$$R_m^n=\frac{1}{2}\left(\phi_{,mn}+\psi_{,mn}\right),\eqno(B.10)$$
$$R_n^4=-\frac{1}{2}\phi_{,n4},\hspace{5mm} R_4^n=\frac{1}{2}\phi_{,n4},
\eqno(B.11)$$
$$R_4^4=\frac{1}{2}\left(\nabla^2\psi-\phi_{,44}-\psi_{,00}\right).\eqno(B.12)$$
\section*{Appendix C: Integration of the Universe Expansion Equation}
The universe expansion was shown to be given by Eq. (3.16), 
$$\frac{dr}{dv}=\tau\left[1+\left(1-\Omega_M\right)r^2/c^2\tau^2\right]^{1/2}.$$
This equation can be integrated exactly by the substitutions 
$$\sin\chi=\alpha r/c\tau;\hspace{5mm}\Omega_M>1\eqno(C.1a)$$
$$\sinh\chi=\beta r/c\tau;\hspace{5mm}\Omega_M<1\eqno(C.1b)$$
where
$$\alpha=\left(\Omega_M-1\right)^{1/2},\hspace{5mm} 
\beta=\left(1-\Omega_M\right)^{1/2}.\eqno(C.2)$$

For the $\Omega_M>1$ case a straightforward calculation using Eq. (C.1a) gives
$$dr=\left(c\tau/\alpha\right)\cos\chi d\chi \eqno(C.3)$$
and the equation of the universe expansion (3.16) yields
$$d\chi=\left(\alpha/c\right) dv.\eqno(C.4a)$$
The integration of this equation gives
$$\chi=\left(\alpha/c\right) v+\mbox{\rm const}.\eqno(C.5a)$$
The constant can be determined using Eq. (C.1a). At $\chi=0$, we have 
$r=0$ and $v=0$, thus
$$\chi=\left(\alpha/c\right) v,\eqno(C.6a)$$
or, in terms of the distance, using (C.1a) again,
$$r\left(v\right)=\left(c\tau/\alpha\right)\sin\alpha v/c;\hspace{5mm}
\alpha=\left(\Omega_M-1\right)^{1/2}.\eqno(C.7a)$$
This is obviously a decelerating expansion.

For $\Omega_M<1$, using Eq. (C.1b), a similar calculation yields for the
universe expansion (3.16)
$$d\chi=\left(\beta/c\right) dv,\eqno(C.4b)$$
thus 
$$\chi=\left(\beta/c\right) v+\mbox{\rm const}.\eqno(C.5b)$$
Using the same initial conditions as above then gives
$$\chi=\left(\beta/c\right)v\eqno(C.6b)$$
and in terms of distances,
$$r\left(v\right)=\left(c\tau/\beta\right)\sinh\beta v/c;\hspace{5mm}
\beta=\left(1-\Omega_M\right)^{1/2}.\eqno(C.7b)$$
This is now an accelerating expansion.

For $\Omega_M=1$ we have, from Eq. (3.16),
$$d^2r/dv^2=0.\eqno(C.4c)$$
The solution is, of course,
$$r\left(v\right)=\tau v.\eqno(C.7c)$$
This is a constant expansion.
\section*{Acknowledgements}
It is a pleasure to thank Julia Goldbaum for calculating all needed
geometrical expressions appearing in the paper, and Tanya Kuzmenko for 
useful discussions.

\end{document}